\documentclass[%
reprint,
superscriptaddress,
longbibliography,
amsmath,amssymb,
prb,
]{revtex4-2}

\usepackage[version=4]{mhchem}
\setcitestyle{numbers,square}
\usepackage{subfigure}
\usepackage{booktabs}
\usepackage{tabularx}
\usepackage{physics}
\usepackage{graphicx}
\usepackage{dcolumn}
\usepackage{bm}
\usepackage{xcolor}
\usepackage{float}
\usepackage{hyperref}
\usepackage{mathtools}
\usepackage[normalem]{ulem}

\hypersetup{
    colorlinks,%
    citecolor=blue,%
    linkcolor=blue,%
    urlcolor=blue
}

\newcommand{\orcid}[1]{\href{https://orcid.org/#1}{\includegraphics[width=8pt]{orcid.pdf}}}

\newcommand{\gy}{GY$^{\text{(1D)}}$}
\newcommand{\gyd}{GYD$^{\text{(1D)}}$}

\usepackage{hyperref}
\hypersetup{
    colorlinks,%
    citecolor=blue,%
    linkcolor=blue,%
    urlcolor=blue
}

\begin{document} 

\title{Self-assembly and Electronic Properties of Graphyne and Graphdiyne Molecular Wires on Metallic Surfaces}

\author{Victor M. S. da Conceição}
\email{manoelsoares1652@gmail.com}
\affiliation{Instituto de F\'{\i}sica,  Universidade Federal de Uberlândia, Av. João Naves de Ávila, 2121 - MG, 38400-902, Brazil}

\author{Dominike Pacine}
\email{dominike@iftm.edu.br}
\affiliation{Instituto Federal de Educação, Ciência e Tecnologia do Triângulo Mineiro, Campus Uberlândia, C.P. 1020, 38064-190, MG,
Brazil}

\author{Juliana M. Morbec}
\email{j.morbec@keele.ac.uk}
\affiliation{School of Engineering and Physical Sciences, Keele University, Keele ST5 5BG, United Kingdom}

\author{Roberto H. Miwa}
\email{hiroki@ufu.br}
\affiliation{Instituto de F\'{\i}sica,  Universidade Federal de Uberlândia, Av. João Naves de Ávila, 2121 - MG, 38400-902, Brazil}

\date{\today}

\begin{abstract}
{\centering\large\bfseries Abstract\par}
Molecular self-assembly on solid surfaces has been the subject of extensive research, motivated by both fundamental and technological interests. On the fundamental side, these studies seek to elucidate the mechanisms governing molecular self-assembly and the resulting surface structures. From an applied perspective, they provide a route toward controlling surface reactions and engineering molecular electronic devices. Here, based on first-principles density functional theory calculations, we present a comprehensive study of self-assembled molecular wires (MWs), composed of graphyne ({\gy}) and graphdiyne ({\gyd}) like structures, adsorbed on Au(111), Ag(111), and Al(111) surfaces. Our total-energy calculations reveals that non-aligned MW arrays are energetically preferred on all three metal substrates. The {\gy} and {\gyd} molecular wires interact with the metal surfaces through van der Waals (vdW) forces, where their molecular orbitals do not contribute to the formation of metallic interface states. Simulated x-ray photoelectron spectroscopy (XPS) spectra reveal that the C-$1s$ spectral features of the molecular wires are largely preserved upon adsorption, while the absolute binding energies undergo a substantial downshift that is nearly independent of the metal substrate, indicating that metallic screening effects dominates the adsorption-induced core-level shifts. Electronic band-structure calculations further show that the semiconducting character of the molecular wires is retained, resulting in vdW metal–semiconductor heterostructures in which the semiconducting component consists of one-dimensional semiconducting channels. These findings demonstrate that self-assembled graphyne- and graphdiyne-based molecular wires on metal surfaces provide a promising platform for the realization of low-dimensional molecular electronic devices.
\end{abstract}

\maketitle
\newpage
\section{Introduction}
Molecular self-assembly on solid surfaces represents a damage-free bottom-up approach to fabricate ordered low-dimensional nanostructures with atomic precision \cite{nanoscale_cartoceti2026,advanced_materials_Yu_2019,Accounts_of_Chemical_Research_Jia_2017}. Integrating flat organic layers without direct chemical bonding yields van der Waals (vdW) metal–semiconductor junctions that remain free of chemical disorder and interface-defect-induced Fermi-level pinning \cite{nature_Liu2018, acs_applied_elec_materialsLee_2023,Zhang_2026_acs_nano}. Among organic platforms, carbon-based molecular architectures such as Graphyne (GY) and graphdiyne (GYD) are promising members of the family of atomically thin, planar carbon allotropes that contain $sp$-hybridised carbon atoms incorporated into $sp^2$-hybridised carbon frameworks. This mixed hybridisation leads to structural flexibility and a variety of polymorphs, enabling their electronic properties to be tuned through the arrangement of carbon-carbon triple bonds, making them particularly attractive for electronic and optoelectronic applications \cite{Li2014}.

Moreover, the presence of acetylenic and diacetylenic linkages provides chemically active sites for surface-assisted self-assembly, facilitating the formation of extended one-dimensional structures (molecular wires) with tunable electronic properties. Theoretical studies \cite{ZHU201557} have shown that the electronic properties of such C$\equiv$C-containing wires can be tailored by controlling the ratio and arrangement of acetylenic units, highlighting the potential of GY- and GYD-based molecular wires for nanoscale electronic applications.

In this context, metallic surfaces serve a crucial dual function: acting as physical templates that guide precursor alignment while simultaneously catalyzing low-temperature chemical coupling. Recent experimental studies have demonstrated the successful on-surface synthesis of GYD molecular wires on Au  \cite{10.1039/d5nr01968k,Au_ullmannStone2022}  and Ag \cite{Ag_ullmannJudd2017} surfaces, showing that substrate orientation strongly influences the efficiency of the organometallic-to-covalent transition during Ullmann coupling. Notably, the room-temperature deposition of 1,3,5-tris(bromoethynyl)benzene (tBEB) on (111)-oriented noble metal surfaces has emerged as a robust route for constructing extended $sp\text{-}sp^2$ nanostructures. This strategy was demonstrated on Au(111) to fabricate graphdiyne-like carbon nanonetworks \cite{rabia_202_acs_applied_nano_mat}, as well as on Ag(111), where D'Agosta et al. \cite{DAgosta-2025_small} obtained a highly crystalline, long-range ordered 2D organometallic lattice. These findings highlight the critical role of metal surfaces in governing the formation of these one-dimensional systems and motivate a broader investigation of substrate-dependent molecular self-assembly. During synthesis, native metal adatoms (e.g., Au, Ag) coordinate with carbon radicals to stabilize intermediate organometallic networks before thermal annealing yields final covalent C–C frameworks \cite{c-cFan2015,c-c2Sun2016}.

Driven by these insights, in this work, based on first-principles density functional theory calculations, we present a comprehensive study of self-assembled one-dimensional molecular wires (MWs), based on graphyne ({\gy}) and graphdiyne ({\gyd}) like structures, adsorbed on Au(111), Ag(111), and Al(111) surfaces (MW/metal). We show that the molecular wires adsorb on the metal surfaces through long-range van der Waals (vdW) interactions. The weak molecule–surface interaction, together with lateral repulsive interactions between neighboring molecular wires, favors the formation of non-aligned arrays, which is in line with the experimental observations of De Boni \textit{et al.}~\cite{DeBoni2024} for molecular wires on Au(111). The structural and electronic properties of the MW/metal systems are characterized through C-$1s$ x-ray photoelectron spectroscopy (XPS) simulations and electronic band-structure analysis. Our results show that adsorption preserves the semiconducting character of the {\gy} and {\gyd} molecular wires, leading to the formation of van der Waals metal–semiconductor heterostructures in which the semiconducting component consists of one-dimensional molecular-wire channels.

\section{Computational Details}

The calculations were performed within the framework of Density Functional Theory (DFT) calculations. First-principles simulations were performed with the VASP (Vienna Ab-initio Simulation Package) \cite{Kresse1993}. To describe the interactions between core and valence electrons, we employed the Projector Augmented-Wave (PAW) method\cite{Blchl1994}. The electronic exchange–correlation interactions were treated using the Generalized Gradient Approximation proposed by Perdew, Burke, and Ernzerhof (GGA-PBE) functional\cite{Perdew1996}.

For the structural relaxation calculations, a plane-wave basis set with a kinetic energy cutoff of 350 eV was employed. For the remaining electronic property calculations, an energy cutoff of 400 eV was used. Van der Waals interactions were taken into account using the nonlocal vdW-DF functional proposed by Dion et al \cite{Dion2004}. 

Brillouin-zone integrations were carried out using the Monkhorst--Pack method \cite{Monkhorst1976}. Due to the use of supercells, the convergence-tested meshes were $5\times4\times1$ for structural relaxations and $5\times5\times1$ for subsequent calculations. All atoms in the supercells were subjected to full relaxation via the conjugate gradient algorithm, reaching a threshold of residual forces below 0.01 eV/Å.

To model the $n$MW/metal interfaces, commensurate $(3\times 4)$ and $(3\times 5)$ supercells were constructed for GY$^{(1D)}$ and GYD$^{(1D)}$ on $\text{Au}(111)$, $\text{Ag}(111)$, and $\text{Al}(111)$, respectively. Along the lattice parameter $a$, a uniform supercell vector of $15 \text{ \AA}$ was adopted across all substrates, corresponding to three times the relaxed primitive surface parameters ($4.15\text{ \AA}$ for $\text{Au}$~\cite{Toro2018}, $4.14\text{ \AA}$ for $\text{Ag}$~\cite{Gu2008}, and $4.15\text{ \AA}$ for $\text{Al}$~\cite{Straumanis1971}), introducing a negligible strain ($< 0.5\%$). Along the periodic wire axis ($b$), lattice matching was achieved by accommodating a minor axial strain strictly within the nanostructures ($\sim 2\%$ for GY$^{(1D)}$ and $\sim 4\%$ for GYD$^{(1D)}$), keeping the underlying metal substrates fully relaxed ($b = 2.87\text{ \AA}$, $2.88\text{ \AA}$, and $2.86\text{ \AA}$ for $\text{Au}$, $\text{Ag}$, and $\text{Al}$, respectively).

Within density functional theory calculations, the core-level binding energies were obtained using the \(\Delta\)SCF approach \cite{GarcaGil2012}. In this method, two distinct self-consistent calculations are performed for each carbon atom: (i) a reference calculation in which all electrons, including the core-level electrons, are explicitly considered, yielding the total energy \(E(n_{c})\); and (ii) an excited-state calculation in which one electron is removed from the \(1s\) core level, resulting in the total energy \(E(n_{c}-1)\). The core-level binding energy \(E_{CL}\) is then defined as

\begin{equation}
E_{CL} = E(n_{c}-1) - E(n_{c}).
\label{eq:core_level_shift_dft}
\end{equation}

In order to establish a common reference for all calculated systems, a methane molecule (\(\mathrm{CH}_{4}\)) was introduced into the simulation cell at a sufficiently large distance from the MWs and substrates, ensuring the absence of electronic interaction between the reference molecule and the studied systems \cite{porcelli2025photoemission}. The \(1s\) core-level binding energy of the carbon atom in the \(\mathrm{CH}_{4}\) molecule was then calculated using the same \(\Delta\)SCF procedure described above, yielding the reference energy \(E_{CL}^{\mathrm{CH_4}}\).

The relative core-level shift for each carbon atom in the MWs was then obtained as

\begin{equation}
\Delta E_{CL} = E_{CL} - E_{CL}^{\mathrm{CH_4}},
\label{eq:relative_cls}
\end{equation}
where \(E_{CL}\) corresponds to the calculated core-level binding energy of a carbon atom in the investigated system and \(E_{CL}^{\mathrm{CH_4}}\) is the corresponding value for the carbon atom in the methane reference molecule.

In order to place the calculated spectra on the experimental energy scale and allow direct comparison with experiment, the experimental C-\(1s\) binding energy (\(BE\)) of methane was added to the calculated relative shifts, such that
\begin{equation}
BE = \Delta E_{CL} + E_{B}^{\mathrm{CH_4,exp}},
\label{eq:absolute_be}
\end{equation}
where \(E_{B}^{\mathrm{CH_4,exp}} = 290.9~\mathrm{eV}\) corresponds to the experimental C-\(1s\) binding energy of the methane molecule \cite{perry1974correlation}.

Once the binding energies were obtained for all carbon atoms, atoms presenting identical or nearly identical energies were grouped together, and a statistical weight proportional to the number of atoms in each group was assigned. 

The XPS spectrum was then convoluted using a pseudo-Voigt function, adopting \(\sigma = 0.38\)~eV as the Gaussian width, \(\gamma = 0.17\)~eV as the Lorentzian width, and a mixing parameter of 0.47 between the two distributions, following the methodology proposed by Ugolotti \textit{et al.}~\cite{Ugolotti2017}.

\section{Results e Discussions}

We first investigate the energetics of the MW/metal systems and subsequently characterize their structural properties through simulations of the C-$1s$ core-level binding energy shifts (CLSs), eq. (2). Finally, we analyze the electronic properties of the adsorbed molecular wires.

\subsection{Energetic and Structural properties}

\begin{figure}[!htb]
    \centering
    \includegraphics[width=1\columnwidth]{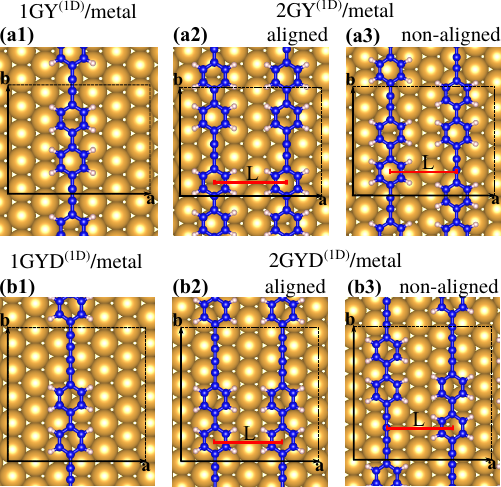} 
    \caption{Schematic top-view representation of the optimized interface models for (a1) single {\gy} and (b1) single {\gyd} molecular wires (1MWs) on metal(111) surfaces. corresponding aligned and non-aligned double-wire configurations ($2\text{MWs}$), respectively. The corresponding aligned double-wire configurations ($2\text{MWs}$) are shown in (a2) for \gy{} and (b2) for \gyd{}, while the corresponding nonaligned configurations are shown in (a3) and (b3), respectively. The inter-wire lateral distance (L) and the supercell boundaries (black dashed lines), defined by the lattice vectors $a$ and $b$, are explicitly indicated.}
    \label{fig:models}
\end{figure} 

Fig.~\ref{fig:models}  illustrates the structural models of  of {\gy} and {\gyd} molecular wires (MWs) adsorbed along the $[11\bar{2}]$ direction of metal (111) surfaces, comparing single [Figs.~\ref{fig:models}(a1)/(b1)], double-aligned [Figs.~\ref{fig:models}(a2)/(b2)], and non-aligned [Figs.~\ref{fig:models}(a3)/(b3)] configurations. Hereafter, single and double aligned configurations are compactly denoted as $n\text{MW/metal}$ ($n = 1, 2$).

We begin by calculating the adsorption energy, $E^a$, of {\gy} and {\gyd} on the metal surfaces, defined as
\begin{equation}
E^a= nE[\mathrm{MW}]+E[\mathrm{metal}] - E[n\mathrm{MW/metal}],
\end{equation}
where $E[n\mathrm{MW/metal}]$ and $E[\mathrm{MW}]$ are the total energies of the adsorbed and free-standing MW systems, respectively, with $n=1$ ($n=2$) for single-wire (double-wire) structures, and $E[\mathrm{metal}]$ is the total energy of the clean metal (111) surface. For each 2MW/metal system, both aligned and non-aligned stacking arrangements were considered. All calculated adsorption energies are summarized in Table \ref{tab:ads}. 
\begin{table}[ht]
\centering
\caption{Adsorption energies of $n${\gy}/metal and $n${\gyd}/metal (in eV/benezene-ring). }
\label{tab:ads}
\begin{ruledtabular}
\begin{tabular}{lcccccc}
\multicolumn{1}{c}{metal} &
\multicolumn{1}{c}{1\gy} &
\multicolumn{2}{c}{2\gy} &
\multicolumn{1}{c}{1\gyd} &
\multicolumn{2}{c}{2\gyd} \\
     \cline{2-2}   \cline{3-4} \cline{5-5} \cline{6-7}
        &       &   alig. & non-alig. & & alig. & non-alig. \\
      
Au(111)$^a$& 0.533& 0.551& 0.566& 0.592&0.627 & 0.638\\
Ag(111) & 0.472 & 0.488& 0.501& 0.562&0.551 & 0.568 \\
Al(111) & 0.513& 0.491& 0.517& 0.568 &0.558 & 0.580 \\ 
\end{tabular}
\end{ruledtabular}
\footnotesize{$^a$ For a benzene molecule on the Au(111) surface we obtained an adsoption energy of 0.55~eV}
\end{table}

Regarding the single-wire systems on Au(111), namely 1{\gy}/Au and 1{\gyd}/Au, we obtain adsorption energies of 0.533 and 0.592~eV per benzene ring, respectively. These values are very close to the adsorption energy calculated for an isolated benzene molecule on Au(111), 0.55~eV, consistent with previous reports in the literature~\cite{Bili2006,Liu2013,Wellendorff2010}. This finding indicates that the benzenoid units of {\gy} and {\gyd} govern the molecule–substrate interaction on Au(111). An analogous scenario is found for Ag(111) and Al(111), where the $sp^2$-hybridized carbon atoms constituting the benzene rings are the primary contributors to the adsorption process.

Upon increasing the molecular coverage, from 1MW/metal to 2MW/metal, the binding strength of both {\gy} and {\gyd} MWs on the metal surfaces increases for the non-aligned configurations. For instance, on Au(111), the adsorption energy changes from 0.533 eV in 1\gy/Au to 0.566 eV/benzene for the corresponding non-aligned 2{\gy}/Au. Moreover, the non-aligned configurations are more stable than the aligned ones by 15 (11) meV per benzene ring for {\gy} ({\gyd}), consistent with the repulsive interaction between phenyl rings reported in Ref.~\cite{DeBoni2024}. 
\begin{table}[ht]
\centering
\caption{Calculated equilibrium molecule-surface separations ($d$ in \AA) for single and double wire configurations on metal substrates.}
\label{tab:geom}
\begin{ruledtabular}
\begin{tabular}{lccccc}
\multicolumn{1}{c}{metal} &
\multicolumn{2}{c}{Single} &
\multicolumn{2}{c}{Double (non-aligned)} \\
\cline{2-3} \cline{4-5}
 & 1\gy & 1\gyd & 2\gy & 2\gyd \\
\hline
Au(111) & 3.58 & 3.59 & 3.68 & 3.69 \\
Ag(111) & 3.69 & 3.67 & 3.75 & 3.74 \\
Al(111) & 3.88 & 3.83 & 3.88 & 3.87 \\
\end{tabular}
\end{ruledtabular}
\end{table}

Beyond the energetic trends, the structural arrangement of the non-aligned double-wire systems adopts the adsorption configuration proposed in in Ref.~\cite{DeBoni2024}. In this context, the chosen transverse supercell dimension ($a = 15 \text{ \AA}$) plays a distinct physical role in $n\text{MW/metal}$ interfaces as a function of wire density. For single-MW setups (one wire per supercell), this spacing fully suppresses lateral inter-wire interactions, simulating isolated molecular wire. In contrast, for the double-wire configurations, comprising two parallel wires per supercell in both aligned and non-aligned arrangements, this periodicity maintains an inter-wire separation of approximately $7.50\text{ \AA}$. As summarized in Table~\ref{tab:geom}, this increased wire density induces a slight outwards shift in the equilibrium molecule-surface separations ($d$), increasing from $\sim 3.58$--$3.88\text{ \AA}$ in single-wire configurations to $\sim 3.68$--$3.88\text{ \AA}$ in double-wire configurations across all three substrates. In particular, the calculated molecule-surface separations ($d$), together with the remarkably low adsorption energies across all $n$MW/metal configurations ($\approx$3 meV/\AA$^{2}$), indicate that the binding of {\gy} and {\gyd} on the metal surfaces is predominantly governed by vdW interactions.

\subsection{Electronic properties}
In the following, we focus exclusively on the non-aligned 2\gy/metal and 2\gyd/metal structures, as they are the most stable configurations and have been experimentally observed on Au(111) \cite{DeBoni2024}.

\subsubsection{Charge transfer analysis}

To gain deeper microscopic insight into the interfacial electronic interactions of $2\text{GY}^{(1\text{D})}$ and $2\text{GYD}^{(1\text{D})}$ on metal (111) surfaces, three-dimensional charge density difference (CDD) distributions were evaluated  based on the following procedure:

\begin{equation}
    \Delta\rho = \rho[\mathrm{MW/metal}] - \rho[\mathrm{MW}] - \rho[\mathrm{metal}],
\end{equation}
where $\rho[\mathrm{MW/metal}]$ represents the total charge density of the combined system, whereas $\rho[\mathrm{MW}]$ and $\rho[\mathrm{metal}]$ denote the charge densities of the isolated components, namely, the pristine MW and the bare metal surface, respectively.

\begin{figure}
    \centering
    \includegraphics[width=\columnwidth]{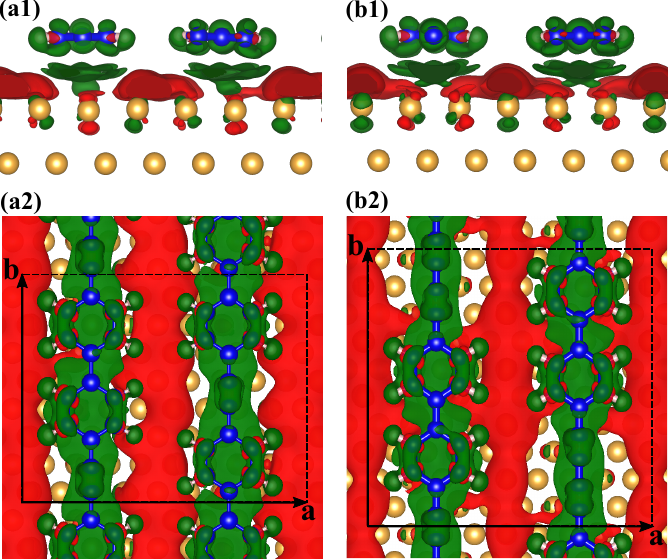} 
    \caption{Net charge transfer for 2\gy/Au and 2\gyd/Au, shown in side and top views in (a1)–(a2) and (b1)–(b2), respectively. All representations consider an isosurface of 1.5 × 10$^{-4}$ e$^{-}$/\AA$^{3}$. Red indicates electron accumulation, whereas green indicates electron depletion.}
    \label{fig:delta_rho}
\end{figure}

For both 2MW/metal systems, the spatial redistribution of charge exhibits electron depletion (green isosurfaces) directly beneath the molecular backbone and electron accumulation (red isosurfaces) extending laterally into the open inter-chain gaps. Figs.~\ref{fig:delta_rho}(a1)--(a2) and \ref{fig:delta_rho}(b1)--(b2) illustrate the charge density differences, $\Delta\rho$, for 2{\gy}/\text{Au} and  2{\gyd}/\text{Au}, respectively. Similar $\Delta\rho$ distribution profiles were obtained for the remaining $2\text{MW/metal}$ systems, with the corresponding data for Ag and Al substrates provided in the Supplemental Material (Fig. S1). Although the organic–metal coupling in these junctions is governed by weak vdW physisorption without chemical bond formation or net charge transfer \cite{pushback_Witte-2005, pushback_Canellas-2020}, Pauli exchange repulsion remains active \cite{pushback_Zojer_2019}. 

The spilling electron density from the clean {metal (111) substrate experiences exchange repulsion due to overlapping closed-shell orbitals of the flat-lying carbon wires. Rather than accumulating purely vertically above the metal surface or being pushed entirely back into the bulk, the dislocated electronic charge is laterally displaced away from the contact zone under the wire backbone into the vacant vacuum channels. This lateral charge migration lowers the kinetic energy penalty associated with wavefunction orthogonalization directly underneath the molecular body, as established for physisorbed molecules on Au(111) contacts \cite{pushback_Zojer_2019,pushback_abad_2011}. 

\subsubsection{Core-level}

In the {2\gy}/metal and {2\gyd}/metal systems, the three inequivalent carbon atoms—namely, the hydrogen-terminated C$_1$, the $sp$-hybridized C$_2$, and the $sp^2$-hybridized C$_3$ atoms—exhibit distinct C-$1s$ core-level binding energies (BEs), which can be probed by x-ray photoelectron spectroscopy (XPS).

Following the procedure described in Sec.~II [Eqs.~(3)–(5)], we begin by calculating the core C-$1s$ core-level BEs of the pristine MWs, obtaining values between 289.5 and 290.7 eV, which are very close to that calculated for an isolated benzene molecule, 289.7~eV, in good agreement with the experimental gas-phase value of 290.36~eV reported in Ref.~\cite{myrseth2002vibrational}. The simulated BE spectra, Figs.~\ref{fig:xps}(a1) and (a2), reveal that the $sp^2$-hybridized C$_3$ atom present the largest BE, followed by C$_1$ (C$_2$) and C$_2$ (C$_1$) in {\gy} ({\gyd}).

\begin{table}[ht] 
\centering 
\caption{Calculated C-$1s$ core-level shifts ($\text{CLS}[C_i] = \text{BE}[C_i] - \text{BE}[C_1]$, in eV) for $2\text{GY}^{(1\text{D})}$ and $2\text{GYD}^{(1\text{D})}$ molecular wires adsorbed on $\text{Au}(111)$, $\text{Ag}(111)$, and $\text{Al}(111)$ surfaces. Shifts are evaluated relative to the hydrogen-terminated $C_1$ carbon binding energy for individual $sp$-hybridized ($C_2$) and $sp^2$-hybridized ($C_3$) species, as well as for the $sp^2$ $C_3$ peak relative to the convoluted $C_1+C_2$ feature.} \label{tab:cls} \begin{ruledtabular} 
\begin{tabular}{lccc} 
\multicolumn{1}{c}{} & 
\multicolumn{2}{c}{CLS[C$_i$]} & 
\multicolumn{1}{c}{CLS[C$_3$]} \\ 
\cline{2-3} \cline{4-4}
\multicolumn{1}{c}{metal} & 
\multicolumn{2}{c}{{2\gy}/metal} & 
\multicolumn{1}{c}{{2\gy}/metal} \\ 
        & C$_2$ & C$_3$ &  C$_3$ \\
Au(111) & -0.10 & 0.34  &  0.37 \\ 
Ag(111) & -0.16 & 0.27  &  0.30 \\ 
Al(111) & -0.23 & 0.36  &  0.38 \\ 
\\ 
\multicolumn{1}{c}{} & 
\multicolumn{2}{c}{{2\gyd}/metal} & 
\multicolumn{1}{c}{{2\gyd}/metal} \\ 
        & C$_2$ & C$_3$ & C$_3$ \\ 
Au(111) & 0.20 & 0.42   &   0.39 \\ 
Ag(111) & 0.06 & 0.48   &   0.46 \\ 
Al(111) & 0.03 & 0.44   &  0.43 \\ 
\end{tabular} 
\end{ruledtabular} 
\vspace{4pt} 
\end{table}

\begin{figure}[!htb]
    \centering
    \includegraphics[width=\columnwidth]{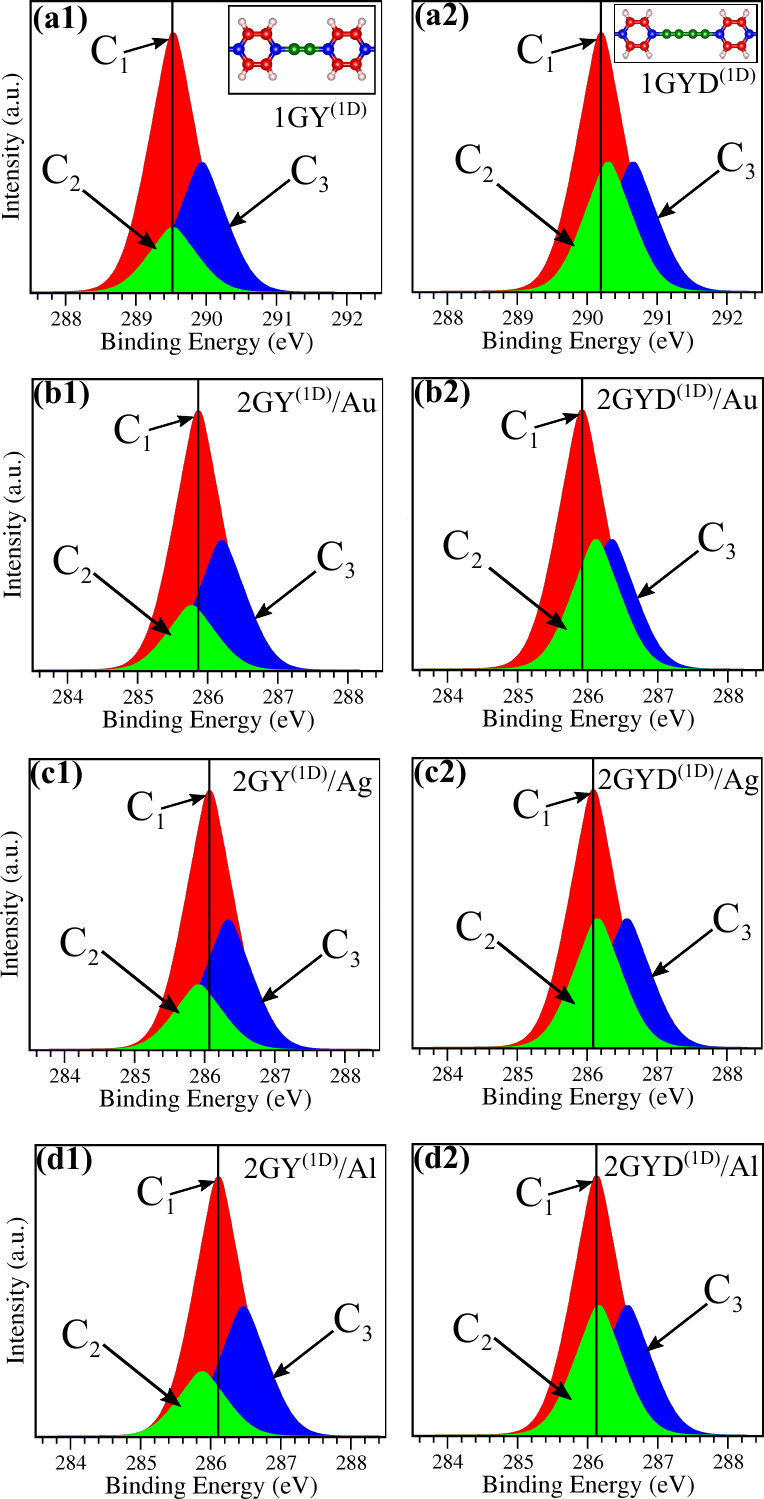}
    \caption{Simulated XPS spectra for isolated molecular wires and adsorbed systems: (a1) free-standing 1\gy and (a2) 1\gyd; (b1)–(b2) double non-aligned 2\gy/Au and 2\gyd/Au; (c1)–(c2) 2\gy/Ag and 2\gyd/Ag; and (d1)–(d2) 2\gy/Al and 2\gyd/Al. C$_1$, C$_2$, and C$_3$ denote the hydrogen-terminated, $sp$-hybridized, and $sp^2$-hybridized carbon atoms, respectively. } 
    \label{fig:xps}
\end{figure}

Figures~\ref{fig:xps}(b)–\ref{fig:xps}(c) show the simulated XPS spectra of the {\gy}/metal and {\gyd}/metal systems. Remarkably, despite the distinct electronic properties of the metal substrates—for example, work functions of 5.27, 4.44, and 4.16 eV for Au(111), Ag(111), and Al(111), respectively—the MW/metal systems exhibit very similar C-$1s$ binding-energy spectra. This behavior indicates that the reduction of the C-$1s$ BEs relative to the corresponding free-standing molecular wires is primarily governed by the interface charge redistribution, $\Delta\rho$, and the associated metallic screening. 

In addition, it is instructive to examine how the core-level shifts (CLSs) depend on both the carbon hybridization and the metal substrate. To this end, we calculate the CLSs of the 2{\gy}/metal and 2{\gyd}/metal systems relative to the binding energy of the hydrogen-terminated carbon atom, C$_1$, BE[C$_1$],
\begin{equation}
    \mathrm{CLS[C_i]}=\mathrm{BE[C_i]} - \mathrm{BE[C_1]},
\end{equation}
with $i$ = 2 and 3. On Au(111), the $sp^2$-hybridized ($sp$-hybridized) carbon atoms of the {\gy} and {\gyd} molecular wires exhibit the largest (smallest) binding-energy splittings relative to the hydrogen-terminated C$_1$ atoms, with CLS[C$_3$] (CLS[C$_2$]) values of 0.34 (-0.10) eV for {\gy} and 0.42 (0.20) eV for {\gyd}, respectively. Table~\ref{tab:cls} summarizes the corresponding CLS[C$_i$] values for the molecular wires adsorbed on the Ag(111) and Al(111) surfaces.


Recently, De Boni \textit{et al.}~\cite{DeBoni2024} reported a comprehensive XPS study of 2{\gy} and 2{\gyd} molecular wires adsorbed on Au(111). For 2{\gy}/Au, they measured binding energies (BEs) of 283.7 and 284.1 eV for the convoluted C$_1$+C$_2$ spectrum and the $sp^2$-hybridized C$_3$ atoms, respectively, corresponding to a core-level shift (CLS) of 0.4 eV. Likewise, for 2{\gyd}/Au, they reported BEs of 283.9 eV for the convoluted C$_1$+C$_2$ spectrum and 284.2 eV for the C$_3$ atoms, yielding a CLS of 0.3 eV.


By convoluting the C$_1$ and C$_2$ spectra of 2{\gy}/Au, we obtain a binding energy of 285.83 eV, while the $sp^2$-hybridized C$_3$ atoms exhibit a BE of 286.20 eV. These values yield a CLS of 0.37 eV between the convoluted C$_1$+C$_2$ spectrum and the C$_3$-$1s$ level; likewise, for 2{\gyd}/Au we obtain a CLS of 0.39 eV. Here, despite the overestimation of the absolute values of BEs, the calculations correctly reproduce the values of CLSs. In Fig.~S2, we present the convoluted spectra of the {\gy}/metal and {\gyd}/metal systems. 

\subsubsection{Electronic band structure}
\label{sec:Electronic_band_structure}

The electronic band structures of the MW/metal systems exhibit metallic character owing to the electronic states of the Au, Ag, and Al substrates. Meanwhile, the projections of the energy bands onto the 2{\gy} and 2{\gyd} molecular wires [Figs.~\ref{fig:bands_projeto_au_al}(a1)--(a3) and \ref{fig:bands_projeto_au_al}(b1)--(b3)] show that the MW-derived states do not contribute to the metallic states at the interface. The band structures of the isolated structures are shown in Fig.~S3.
\begin{figure}[H]
    \centering
   \includegraphics[width=\columnwidth]{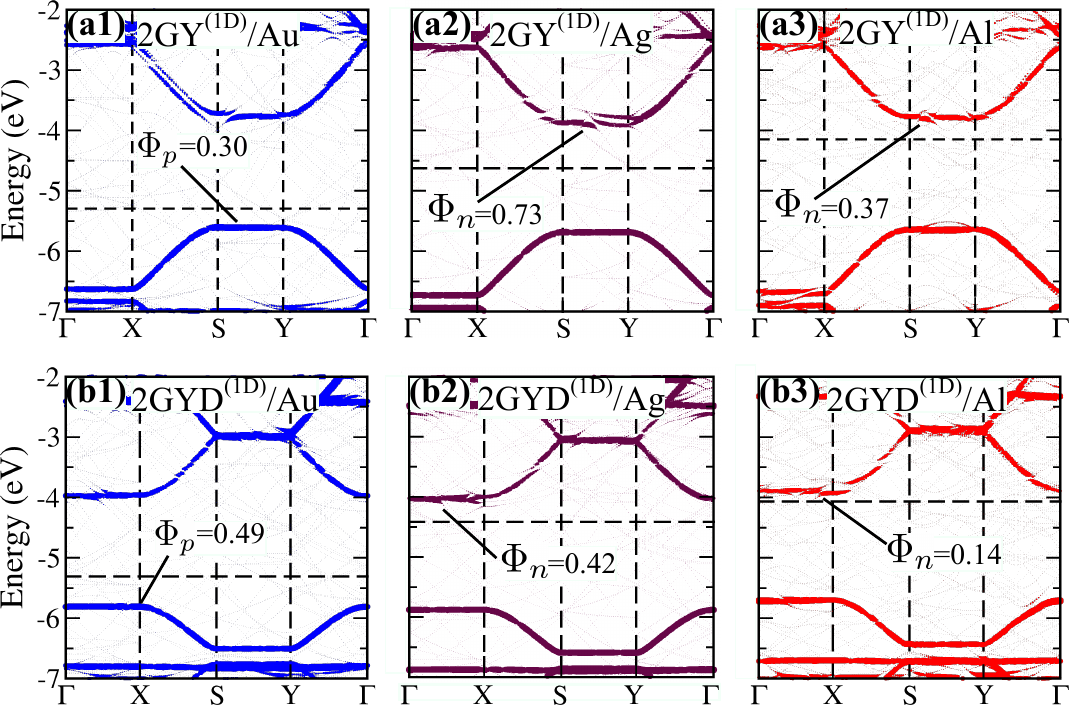} 
    \caption{The band structure of 2\gy, in (a1), (a2) and (a3), on Au, Ag and Al, respectively. Similarly, in (b1), (b2) and (b3) we have the band structures of 2\gyd, on Au, Ag and Al, respectively. The blue and red points indicate the projection of the MW states onto the substrate+MW system bands. The zero was set to the vacuum }
    \label{fig:bands_projeto_au_al}
\end{figure} 
The electronic band structures of the MW/metal systems projected onto the molecular wires are characterized by nearly dispersionless bands along the $\Gamma$X and SY directions in Fig.~\ref{fig:bands_projeto_au_al}, together with dispersive bands along the $\Gamma$Y and XS directions, corresponding to electronic states with wave vectors (k) parallel to the molecular-wire axis. To elucidate the nature of these states, Figs.~\ref{fig:projeção}(a1)--(a2)/(c1)--(c2) present the real-space projections of electronic states with wave vectors perpendicular ($\mathbf{k}_\perp$) to the 2\gy\ and 2\gyd\ molecular wires on Au(111). Correspondingly, the parallel components ($\mathbf{k}_\parallel$) are shown in Figs.~\ref{fig:projeção}(b1)--(b2)/(d1)--(d2). The states associated with $\mathbf{k}_\perp$ are characterized by localized C-$\pi$ orbitals, whereas those associated with $\mathbf{k}_\parallel$ exhibit delocalized C-$\pi$ orbitals extending along the molecular-wire axis.

Notably, adsorption preserves not only the semiconducting character of the molecular wires but also the energy positions of their band edges, with the  Fermi level lying within the band gap of the molecular wires. For instance, in 2{\gy}/Au, the valence-band maximum and conduction-band minimum [Fig.~\ref{fig:bands_projeto_au_al}(a1)] are located approximately 5.6 and 3.8 eV below the vacuum level, respectively. These values are very close to the ionization potential (IP) and electron affinity (EA) of the free-standing 2{\gy} molecular wire, calculated to be 5.4 and 3.6 eV, respectively. Consistently, the work function (WF) of 2{\gy}/Au (5.30 eV) remains essentially unchanged relative to that of the clean Au(111) surface (5.27 eV), in line with the weak electronic coupling between the molecular wire and the substrate. 

Despite the electronic screening of the C-$1s$ core states, the weak molecule–substrate interaction preserves the electronic structure of the molecular wires, such that the MW/metal systems are well described as a superposition of the electronic structures of the isolated constituents. Consequently, the variation in the substrate work function, from 5.27 eV for Au(111) to 4.16 eV for Al(111), leads to the formation of van der Waals metal–semiconductor interfaces, where the semiconducting phase is provided by a layer of non-aligned {\gy} or {\gyd} molecular wires.

\begin{figure}
    \centering
    \includegraphics[width=\columnwidth]{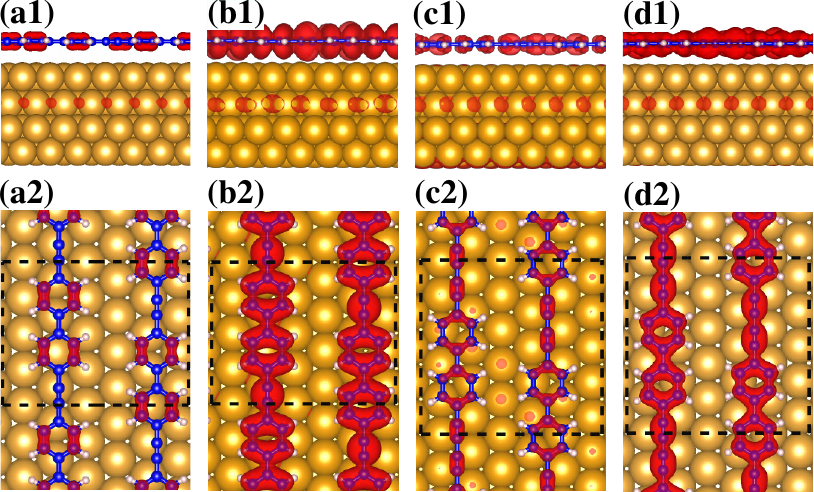} 
    \caption{Side (top row) and top (bottom row) views of the partial charge density projections on Au(111). Projections along $\Gamma\rightarrow X$ are shown for (a1)--(a2) 2\gy/Au and (c1)--(c2) 2\gyd/Au, while projections along $Y\rightarrow\Gamma$ are shown for (b1)--(b2) 2\gy/Au and (d1)--(d2) 2\gyd/Au. An isosurface value of $4\times 10^{-4}$~$e^{-}/\text{\AA}^{3}$ is used in all cases, and black dashed lines indicate supercell boundaries.}
    \label{fig:projeção}
\end{figure}

As shown in Figs.~\ref{fig:bands_projeto_au_al}(a1) and \ref{fig:bands_projeto_au_al}(b1), adsorption of {\gy} and {\gyd} molecular wires on Au(111) gives rise to a $p$-type Schottky contact, whereas adsorption on Al(111) results in an $n$-type Schottky contact [Figs.~\ref{fig:bands_projeto_au_al}(c1) and \ref{fig:bands_projeto_au_al}(c2)]. Owing to the absence of metal-induced gap states at the interface, the Schottky barrier height (SBH) can be estimated directly from the energy separation between the molecular-wire band edges and the Fermi level of the MW/metal system. For a $p$-type contact, the SBH is given by $\Phi_p= E_\mathrm{VBM}-E_\mathrm{F}$, where $E_\mathrm{F}$ is the Fermi level of the MW/metal system and $E_\mathrm{VBM}$ is the valence-band maximum of the molecular wire. Likewise, for an $n$-type contact, $\Phi_n=E_\mathrm{F} - E_\mathrm{CBM}$, where $E_\mathrm{CBM}$ denotes the conduction-band minimum of the molecular wire.

The calculated values of $\Phi_p$ and $\Phi_n$, shown in the insets of Fig.~\ref{fig:bands_projeto_au_al}, indicate that 2{\gy} and 2{\gyd} molecular wires form $p$-type Schottky contacts on Au(111), whereas $n$-type Schottky contacts are obtained on Ag(111) and Al(111). Here, the absence of metal-induced gap states indicates that the Fermi level is not pinned in these MW/metal systems. Consequently, the Schottky barrier height can, in principle, be tuned by external perturbations, such as an external electric field, enabling a transition from a Schottky to an Ohmic contact. 

As an example, a downward (upward) shift of the Fermi level in 2{\gy}/Au (2{\gyd/Al)})  would reduce $\Phi_p$ ($\Phi_n$) and eventually lead to the formation of an Ohmic contact. In this regime, hole (electron) transport would take place along with $\mathbf{k}_\parallel$ along the wire axis.

\section{Conclusion}

In this work, we performed first-principles density functional theory (DFT) calculations to investigate molecular wires (MWs) based on graphyne ({\gy}) and graphdiyne ({\gyd}) adsorbed on Au(111), Ag(111), and Al(111) surfaces. We confirmed that the non-aligned adsorption geometry is energetically preferred, in agreement with the experimental observations of De Boni \textit{et al.}. The calculated adsorption energies of {\gy} and {\gyd} are comparable to that of benzene on Au(111), indicating that the interaction is dominated by van der Waals forces. Consistently, the molecular orbitals of the MWs do not contribute to the formation of metallic interface states.

Simulated C-$1s$ x-ray photoelectron spectra show that the spectral features of the free-standing {\gy} and {\gyd} molecular wires are largely preserved upon adsorption. Although the absolute binding energies are reduced by approximately 3.5 eV for 2{\gy}/metal and 4.1 eV for 2{\gyd}/metal, this reduction is nearly independent of the substrate work function, indicating that it is primarily governed by metallic screening and the associated final-state effects.

The electronic-structure calculations further reveal that the semiconducting character of the molecular wires is retained upon adsorption. The valence- and conduction-band edge states remain delocalized along the molecular-wire axis while exhibiting non overlap in the transverse direction, resulting in 1D electronic confinement along MWs.  Owing to the weak molecule–substrate interaction, the MW/metal systems are well described as van der Waals metal–semiconductor heterostructures. Depending on the substrate work function, the resulting interfaces exhibit either $n$-type or $p$-type Schottky contacts, with Schottky barrier heights ranging from 0.14 eV (2{\gyd}/Al) to 0.30 eV (2{\gy}/Au).

These findings demonstrate that graphyne- and graphdiyne-based molecular wires on metal surfaces constitute a versatile platform for engineering low-dimensional 1D-semiconductor/metal interfaces with tunable electronic properties, opening promising perspectives for molecular electronics and nanoscale device applications.

\begin{acknowledgments}

This work was supported by the Brazilian agencies FAPEMIG, CNPq, and CNPq/INCT (National Institute of Science and Technology on
Materials Informatics, grant n. 371610/2023-0). We would like to acknowledge computing time provided by the National Laboratory for Scientific Computing
(LNCC/MCTI) for providing HPC resources of the SDumont supercomputer.

\end{acknowledgments}

\bibliographystyle{apsrev4-1} 
\bibliography{tex} 
\end{document}


\title{Supporting Information: Self-assembly and Electronic Properties of Graphyne and Graphdiyne Molecular Wires on Metallic Surfaces}

\author{Victor M. S. da Conceição$^1$}
\email{manoelsoares1652@gmail.com}
\author{Roberto H. Miwa$^1$}
\affiliation{$^1$Instituto de F\'{\i}sica,  Universidade Federal de Uberlândia, Uberlândia, 38400-902, MG, Brazil}

\author{Dominike Pacine$^2$}
\affiliation{Instituto Federal de Educação, Ciência e Tecnologia do Triângulo Mineiro, Campus Uberlândia, C.P. 1020, 38064-190, MG,
Brazil}

\author{Juliana M. Morbec$^3$}
\affiliation{School of Engineering and Physical Sciences, Keele University, Keele ST5 5BG, United Kingdom}

\maketitle

\section{}

\begin{figure}[!htb]
    \centering
    \includegraphics[]{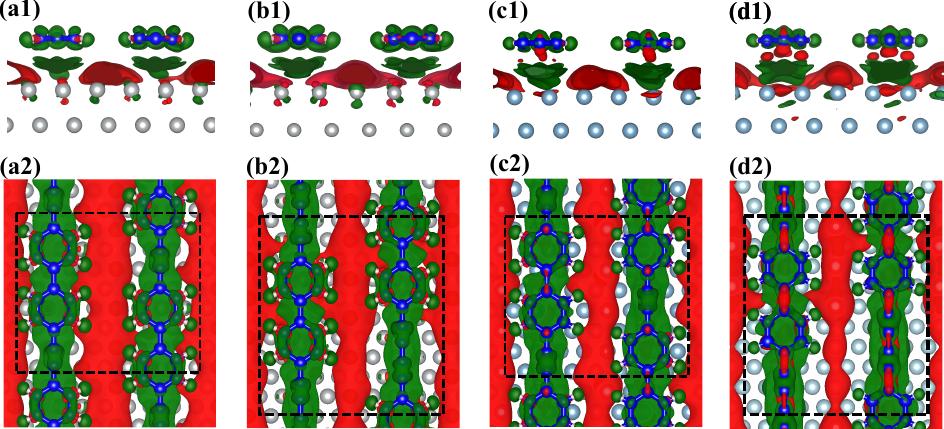}
   \caption{(a1) and (a2) show the side and top views, respectively, of 2\gy on the Ag surface, while (b1) and (b2) show the corresponding views for 2\gyd. Similarly, (c1) and (c2) show the side and top views, respectively, of 2\gy on the Al surface, while (d1) and (d2) show the corresponding views for 2\gyd.}
    \label{fig:delta_rho_ag_al}
\end{figure}

\begin{figure}[!htb]
    \centering
    \includegraphics[]{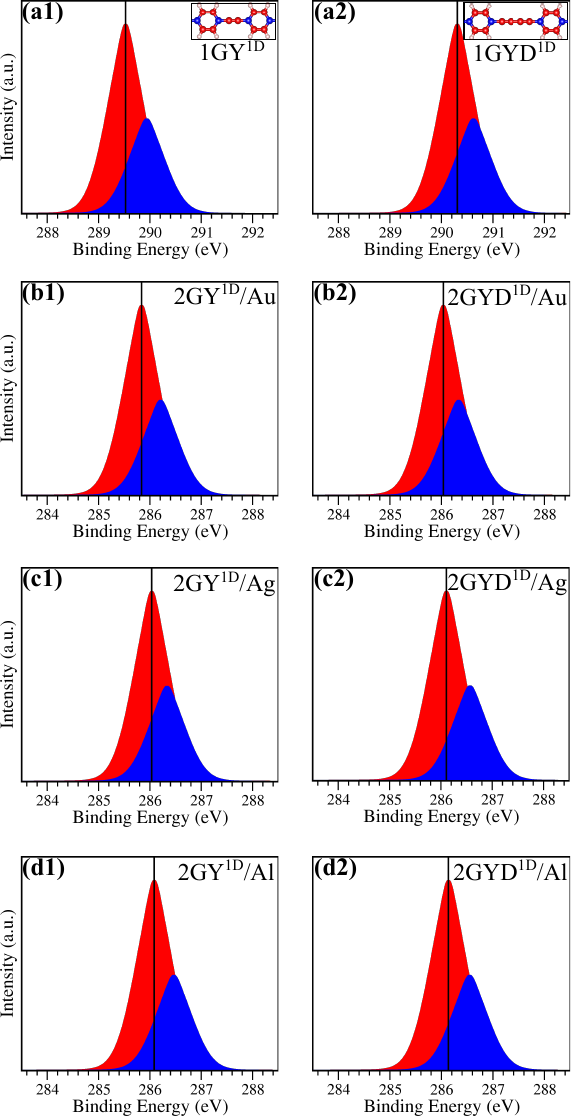}
   \caption{The simulated XPS spectra for isolated molecular wires and adsorbed systems: (a1) freestanding $1\text{GY}^{(1\text{D})}$ and (a2) $1\text{GYD}^{(1\text{D})}$; double non-ligned configurations on metal substrates (b1--b2) $2\text{GY}^{(1\text{D})}/\text{Au}$ and $2\text{GYD}^{(1\text{D})}/\text{Au}$, (c1--c2) $2\text{GY}^{(1\text{D})}/\text{Ag}$ and $2\text{GYD}^{(1\text{D})}/\text{Ag}$, and (d1--d2) $2\text{GY}^{(1\text{D})}/\text{Al}$ and $2\text{GYD}^{(1\text{D})}/\text{Al}$. The simulated spectra are deconvoluted into contributions from the hydrogen-terminated $C_1$ + $sp$-hybridized $C_2$ carbon species (red curve) and the $sp^2$-hybridized $C_3$ benzene ring atoms (blue curve). All energy scales are aligned relative to the methane ($\text{CH}_4$) gas-phase reference.}
    \label{fig:xps_2}
\end{figure}

\begin{figure}[!htb]
    \centering
    \includegraphics[width=\columnwidth]{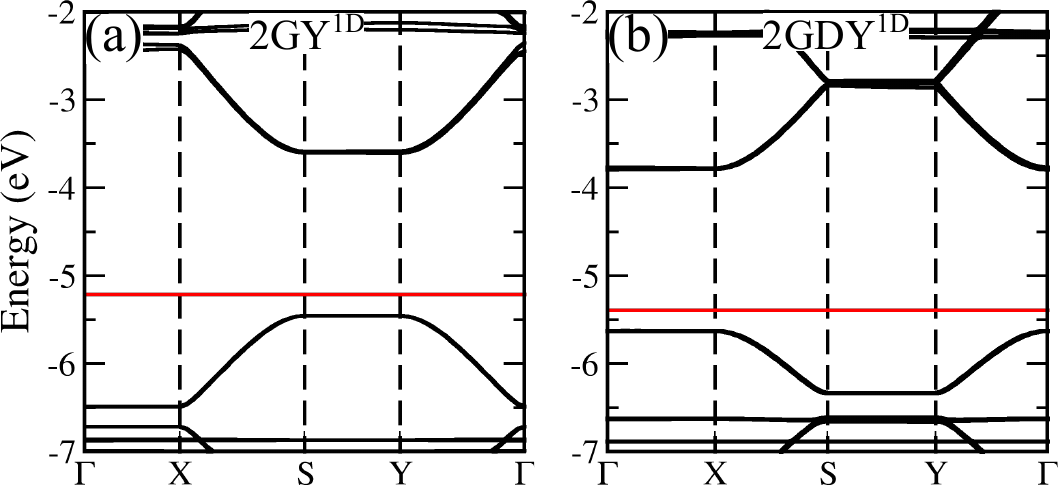} 
    \caption{The band structure of freestanding 2\gy\ system is shown in (a). In (b) shows the band structure of the freestanding 2\gyd\ system. The energy zero is set at the vacuum level, and the solid red horizontal line indicates the Fermi level.} 
    \label{fig:bands}
\end{figure}

